# Use of Augmented Reality in Human Wayfinding: A Systematic Review

Zhiwen Qiu[1,2], Armin Mostafavi[1], Saleh Kalantari[1]

## Abstract:

Augmented reality (AR) technology has emerged as a promising solution to assist with wayfinding difficulties, bridging the gap between obtaining navigational assistance and maintaining an awareness of one's real-world surroundings. In this article, we present a systematic review of research literature related to AR navigation technologies. An in-depth analysis of 65 salient studies was conducted to address four main research topics: (1) the current state-of-the-art of AR navigational assistance technologies, (2) user experiences of interacting with these technologies, (3) the effect of AR on human wayfinding performance, and (4) the impacts of AR for human navigational cognition. One of the notable study findings is that a robust body of literature exists demonstrating that AR can help decrease cognitive load and improve the development of cognitive maps, compared to traditional guidance modalities. However, findings about wayfinding performance and user experience were more mixed, with AR have little demonstrated impact in improving outdoor navigational performance, and with some information modalities shown to be overly distracting and ineffective. The article discusses these nuances in detail, while overall lending support to the conclusion that AR has a great potential to enhance wayfinding by providing enriched navigational cues, interactive experiences, and improved situational awareness.

[1]Human Centered Design, Cornell University
[2]Author's Address: Zhiwen Qiu, Cornell University, Ithaca, NY, USA, zq76@cornell.edu

**1. Introduction**

Wayfinding is a complex cognitive process that requires assimilating spatial information from the environment, applying problem-solving skills, and making and executing decisions to reach a destination (Passini, 1996). This can be challenging for many people, especially in complex modern environments such as hospitals, airports, and large office buildings, which have been associated with wayfinding struggles that may potentially lead to negative practical and psychological outcomes (Arthur & Passini, 1992; Carpman & Grant, 2002). For example, wayfinding problems in healthcare settings can lead to confusion, anxiety, frustration, stress, elevated blood pressure, and fatigue, as well as to missed appointments (Schmitz, 1997; Shumaker & Reizenstein, 1982). To address these problems, technologies such as Augmented Reality (AR) have been introduced to aid wayfinding behaviors and assist in developing internal cognitive maps (Zhang et al., 2021). AR overlays computer-generated navigation instructions onto the real environment using mobile real-time camera screens or head-mounted displays (HMDs), allowing users to remain engaged with their surroundings by not having to split their attention away to a separate device for navigational guidance (Azuma, 1997; Rusch et al., 2013).

The rapid development of this technology has been associated with an explosion of AR wayfinding research, with a substantial increase in the annual number of publications over the past decade. A great deal of this research has been focused on technical challenges, such as the need for reliable localization and tracking methods to synchronize the AR displays with the user's real-world viewpoint. Such methods may involve Simultaneous Localization and Mapping (SLAM), the use of automated registration data to create large-scale 3D model of the surrounding environment (Radanovic et al., 2023), or hybrid tracking that integrates multiple data sources such as Global Positioning System (GPS) and Inertia Measurement Units (IMUs) consisting of accelerometers, gyroscopes, and magnetometers (Delgado et al., 2020). Equally important is the development of robust pathfinding algorithms capable of devising optimal routes for users based on their current locations and the environmental conditions. AR interaction modalities have also diversified significantly, incorporating gesture recognition, touch interaction, and voice commands, among others.

In addition to these technical challenges, however, it is important to conduct robust research on how users interact with AR on a cognitive and experiential level. Such work is crucial to inform the design of more intuitive and user-friendly AR navigational assistance systems, which can enhance their effectiveness and user satisfaction. The primary objective of the current literature review was to evaluate and summarize recent research on AR-assisted wayfinding as related to user experience (UX), wayfinding performance, and cognitive impacts. To guide the review, we formulated the following four research questions:

- **RQ1.** What is the current state-of-the-art of AR navigational assistance technologies?
- **RQ2.** What consensus knowledge has emerged about factors related to the user experience and acceptance of these technologies?
- **RQ3.** Does the research literature consistently show an effect of AR navigational assistance on human wayfinding performance?
- **RQ4.** What are the impacts of AR navigational assistance on human perception, decision-making, behavior, and cognition?

The initial Background section of the paper provides an overview of AR and its applications in the context of human wayfinding. We then present the methods for the current literature review, followed by a comprehensive summary of our findings. The Discussion section analyzes and interprets these results in light of the four research questions outlined above. Finally, we conclude by identifying challenges, trends, and recommendations for future directions in research and practice.

## 2. Background

Augmented reality is a technology that overlays computer-generated virtual content onto the real world, so that the virtual information is aligned with real-world objects and can be viewed in real time (Azuma, 1997). The two main approaches to AR include the use of a screen with a real-time video feed, or the use of transparent head-mounted displays ("smart glasses") to present information. For many years, such AR technologies were confined to research laboratories and military training applications. The advent of tracking systems for mobile phones in the early 2000s marked a pivotal point in the evolution of AR systems, eventually leading to the maturation of the technologies and their assimilation into everyday devices (Arth et al., 2015). Wayfinding assistance was one of the first and most central uses of AR technologies, with early adopters relying on smartphones and designated visual markers to enhance indoor navigation. For example, Möller and colleagues (2014) demonstrated a localization technique using image recognition, in which camera-captured images that matched those in an image library would trigger navigational instructions such as arrows and distance-to-destination superimposed onto the phone's screen. These approaches ultimately evolved into complex AR head-mounted displays that show path visualizations, anchored signs, and layout maps while allowing users to interact with virtual objects via hand-gestures (Oh et al., 2017; San Martin & Kildal, 2021).

In general, AR navigational systems are grounded in three core components. First, spatial localization is used to determine the user's current location in the environment and often their specific field of view. This can be done through some combination of visual markers, IMUs, Bluetooth beacons, GPS, or image-recognition technologies. Some used spatial mapping for more accurate and robust tracking by producing a three-dimensional representation of the environment, often by using cameras or laser scanners to determine the location of surfaces and generate a virtual model (Weinmann et al., 2021). Second, algorithmic path-generation is used to determine an optimal route from the user's current location to their intended destination, by applying pathing algorithms such as Dijkstra (Fan & Shi, 2010) or A* (Duchoň et al., 2014). Finally, the AR system presents visual or auditory instructions, which may include a combination of maps, action-signs, turn-by-turn directions, and/or supplementary point-of-interest information (Lu et al., 2021; Oliveira de Araujo et al., 2019; Zhang et al., 2021; Zhao et al., 2020).

## 2.1. Defining and Measuring User Experiences of AR

User experience in AR includes elements such as emotional responses, interface preferences, physical interactions, and psychological reactions (Arifin et al., 2018; Dirin & Laine, 2018). This experience is an evolving process, unfolding in accordance with the technology's features as the user interacts with the AR device throughout the navigation process. In general, usability refers to the extent to which a technology is perceived as helpful for achieving one's goals effectively, efficiently, and enjoyably (ISO, 1998; Brooke, 1996; Davis, 1989). Aspects of user experience that have been described as particularly salient for AR technologies include: (a) the ability to quickly and easily familiarize oneself with the technology during first contact, (b) the ability to use the technology while moving through a real-world environment without excessive distraction, and (c) the ability to remember the system's functions across multiple sessions (Punchoojit & Hongwarittorrn, 2017). User experience can also be enhanced by accommodating diverse preferences in regard to AR interaction modalities or interfaces. The extent to which people accept and embrace novel technologies is closely tied to such usability factors, and can be empirically measured via scales such as the Technology Acceptance Model (TAM) (tom Dieck & Jung, 2018).

A related construct is users' "attitude" or disposition toward the technology, which focuses on responses such as trust, confidence, and satisfaction (Hassenzahl, 2018). Trust may be affected by the predictability of the system's behavior, prior experiences indicating that the system is dependable in achieving the user's goal, and general perceptions of related technologies (Muir, 1994). In addition to designing a well-performing system, trust can be improved through clear instructional materials and by emphasizing robust security measures to protect user data. This is highly important since a lack of trust can have a significant impact on users' acceptance of AR systems (Corritore, Kracher & Wiedenbeck, 2003). Attitudes toward AR technologies can also be improved by designs that incite positive emotions (pleasantness, excitement) and that minimize negative emotions (frustration, confusion) (Norman, 2004).

In AR navigational assistance systems, users' perception of the informational and physical environment is crucial. Perception refers to the identification and interpretation of sensory information such as depth of field when it comes to the overlayed positioning of virtual objects on the physical world (Schacter, Gilbert & Wegner, 2011; Kim et al., 2018). Achieving fluid visual perception in AR requires very precise localization algorithms as well as an effective design of the virtual elements. Misalignment between the virtual information and the real world can lead to navigational errors and will severely erode the UX aspects of the technology. In contrast, well-aligned perceptual elements can enhance users' situational awareness, promoting a comfortable understanding of where one is, where one is going, and what is happening in the surroundings (Endsley, 1995).

The concepts of immersion and presence are also useful for analyzing users' interactions with AR technologies. Immersion in AR refers to the extent to which a user feels fully engaged or absorbed in the virtual content, perceiving it as a central focus of their experience (Slater & Wilbur, 1997). Fidelity, spatial stability, interactivity, and coherence with the real-world context are all important for achieving AR immersion (Bowman & McMahan, 2007; Slater, 1999). Presence describes the subjective sensation of "being" within the virtual environment rather than just engaging with it. This experience, in which the distinction between the physical world and the virtual AR components fades from users' awareness, is affected by the fidelity of the technological components but is also profoundly shaped by the realism and emotional engagement of the virtual content (Slater & Steed, 2000). Presence has been consistently correlated with higher satisfaction levels in multiple types of virtual environments (Slater & Wilbur, 1997; Bulu, 2012).

## 2.2. Foundational Concepts in Wayfinding Cognition

Wayfinding cognition refers to the complex mental processes involved in human navigation through physical spaces, including understanding, planning, and executing a route from one location to another (Montello, 2005). It involves both conscious and unconscious thought, memory, perception, and spatial problem-solving, and encompasses various forms of spatial memory. Spatial memories themselves can be categorized into three types: landmark knowledge (concerning specific, easily recognizable points in the environment); route knowledge (sequences connecting objects or places); and survey knowledge (an understanding of distances and spatial relationships throughout the environment) (O'Keefe & Nadel, 1979; Siegel and White, 1975; McNamara, 2013). All three types of spatial memory are relevant for the dynamic, ongoing formation of cognitive maps, the mental representations that allow individuals to visualize and reason about spatial relations (Tolman, 1948). Recent studies of navigation have added the concept of "cognitive graphs," which focus on idiothetic information such as path lengths and junction angles as a way of remembering routes (Warren et al., 2017). Current research seems to indicate that human wayfinders rely on both cognitive maps and cognitive graphs, in varying and individualistic combinations, when completing wayfinding activities (Peer et al., 2021).

Cognitive Load Theory posits that human working memory, essential for learning and problem-solving, is limited in capacity and requires significant energy to maintain (Buchner, Buntins & Kerres, 2021). This concept has emerged as a central topic in AR navigation research, under the assumption that AR navigational tools can help to reduce cognitive load, especially when compared against other kinds of navigational devices. The theory stipulates that cognitive load has an intrinsic component, related to task complexity; an extraneous component, tied to the features and design of the environment; and a germane component, concerned with the construction of mental schemas (Kalyuga, 2011; Sweller, van Merriënboer & Paas, 1998). The often-stated goal of AR navigational technologies is to reduce the extraneous load and thus, indirectly, the germane load (Paas & van Merriënboer, 2020). Some prior studies have shown that wearable AR displays can in fact reduce cognitive load during wayfinding and spatial assembly tasks (Deshpande & Kim, 2018; McKendrick et al., 2016). However, the specific design of the AR system is likely relevant in this context, as an excess of information or a poorly designed interface could potentially add to, rather than reduce, the user's cognitive load.

## 3. Methods

We conducted a systematic literature review to identify, evaluate, and synthesize the existing body of relevant scholarship (Fink, 2019; Xiao & Watson, 2019). The PRISMA (Preferred Reporting Items for Systematic Reviews and Meta-analyses) method was used to guide the screening and analysis (Liberati et al., 2009).

**3.1. Search Strategy**

The search of published academic studies was carried out in February of 2023, and the review protocol was registered with Open Science Framework (OSF) in the same month. To help ensure that we evaluated a wide-ranging and varied array of published articles, four databases were chosen, including Scopus, Web of Science, IEEE Xplore, and the ACM Digital Library. All of these databases cover a broad range of scientific publications, rendering them highly suitable for our project's goals.

The following terms were used to search article titles, abstracts, keywords, and subject headings: *("Augment* Reality" OR "Mixed Reality" OR "HoloLens") AND ("Pedestrian Navigat*" OR "Indoor Navigat*" OR "Outdoor Navigat*" OR "Wayfinding" OR "Way-Finding" OR "Pathfinding" OR "Path-Finding" OR "Spatial Cognition").* This string was intended to cover variations of the relevant words (e.g., navigation vs. navigating) and closely related concepts that are often used interchangeably (e.g., augmented reality vs. mixed reality), while still keeping the search narrow enough to exclude the much broader literature on virtual reality. The term HoloLens (Microsoft Corporation, 2023) was included since it is the most commonly used brand of AR headset. We consulted with an evidence synthesis librarian at our university to carefully formulate and apply this search string so that the review would achieve an effective scope. The results included 912 items from Scopus, 656 from Web of Science, 424 from IEEE Xplore, and 143 from the ACM Digital Library.

**3.2. Eligibility Criteria and Screening of Studies**

To ensure relevance, rigor, and coherence in the compilation of data, we established clear inclusion and exclusion criteria. To be *included* in the review, an item was required to satisfy all of the following conditions:

(i1) Published in peer-reviewed journals or peer-reviewed conference proceedings. This criterion helps to ensure the quality and credibility of the research.

(i2) Based on empirical study. We only considered articles that incorporated observations and measurements of empirical phenomena. Opinion articles and those based purely on theory were not included.

(i3) Addressed the use of AR technologies with mobile devices and HMDs for pedestrian navigation (either indoor or outdoor).

Items were *excluded* if they fell into one or more of the following categories:

(x1) Dissertations, books, book chapters, overview articles, editorials, conference posters, workshops, and literature reviews (scoping, systematic, meta-analysis, integrative reviews) were excluded from the study. These types of publications were excluded because they generally present overviews or summaries of research fields, rather than providing new empirical data.

(x2) Items not written in English were excluded due to the potential for language barriers to negatively affect the accurate interpretation and analysis of the research.

(x3) Papers that focused primarily on virtual reality, rather than AR or mixed reality, were excluded. Our review covered the use of AR to support real-world wayfinding and did not address the issue of navigating in purely virtual environments.

(x4) Papers that focused primarily on non-pedestrian navigation, for example navigation while driving an automobile, were excluded. Our review only covered the use of AR in pedestrian wayfinding.

Figure 1 presents a PRISMA flowchart for our article selection process. We first used the Covidence software platform (Veritas Health Innovation, 2023) to automatically remove duplicate studies. The remaining 884 articles were screened by abstract in accordance with our inclusion criteria. Most of the studies removed in this phase did not address pedestrian navigation—we found that a large number of papers on self-driving automobiles had slipped through our search criteria, as well as a large number of papers on general AR development that did not focus on navigation at all. After the abstract screening, 214 items were retrieved for full-text eligibility evaluation. In this phase, the majority of the removed studies did not use empirical research methods, with additional removals for not addressing pedestrian navigation and various other breaches of the inclusion/exclusion criteria. At the end of the multi-stage screening process, 65 studies remained. A short description of each included study is given in Table 1.

**Figure 1.** PRISMA Flowchart Showing the Selection of Studies

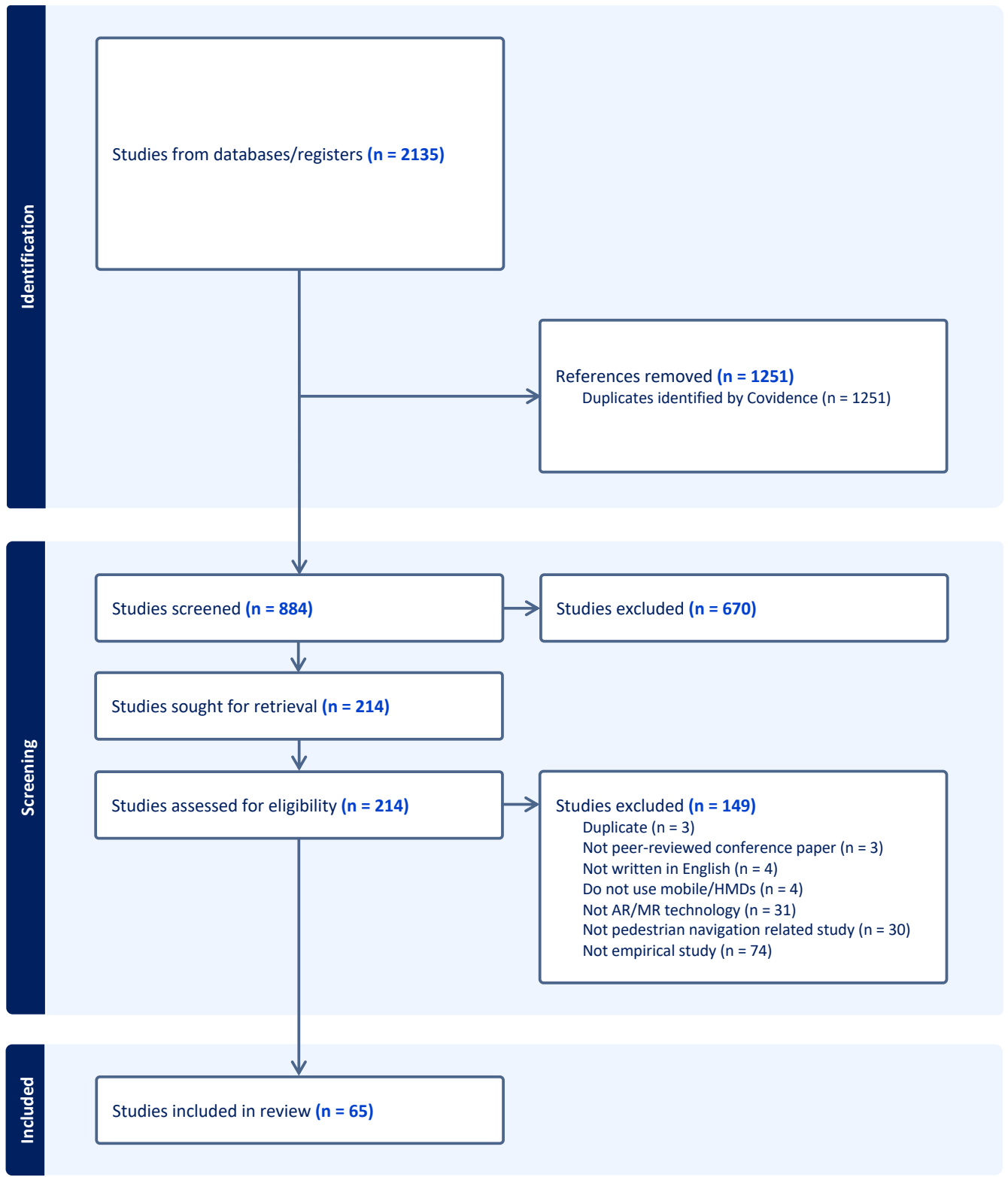

**Table 1.** Summary Description of Studies Included in the Systematic Review

|   | **Paper** | **Topic of Study** |
|---|-----------|---------------------|
| 1 | Stigall et al., 2019 | Use of Microsoft HoloLens for navigational assistance in evacuation scenarios. |
| 2 | Rochadiani et al., 2022 | Navigating through a shopping mall environment using augmented reality. |
| 3 | Ahmad et al., 2005 | Gender differences in navigation and wayfinding using mobile augmented reality. |
| 4 | Kim et al., 2015 | User experiences with an augmented reality-enabled wayfinding system in complex environments. |
| 5 | Smith et al., 2017 | Using augmented reality to improve navigation skills in postsecondary students with intellectual disability. |
| 6 | Ping et al., 2020 | Applying a mobile AR guide system to enhance user experience in cultural heritage sites. |
| 7 | Hou & Tang, 2020 | Contrast and parameter research for an augmented reality indoor navigation scheme. |
| 8 | Arntz et al., 2020 | Navigating a heavy industry environment using augmented reality, comparing two types of AR designs. |
| 9 | Romli et al., 2020 | Mobile augmented reality marker-based approach for indoor library navigation. |
| 10 | Drewlow et al., 2022 | Use of augmented reality to improve navigation in a hospital environment. |
| 11 | Sheoprashad & Defreitas, 2022 | Investigating indoor navigational experiences with a mobile augmented reality prototype versus conventional methods. |
| 12 | Wakchaure et al., 2022 | Indoor navigation system for public evacuation in emergency situations. |
| 13 | Yunardi et al., 2022 | Design and development of an object-detection system in AR for indoor navigation. |
| 14 | Sharin et al., 2023 | Combining a step-counting technique with augmented reality for a mobile-based indoor localization. |
| 15 | Kluge & Asche, 2012 | Validating a smartphone-based pedestrian navigation system prototype with eye-tracking. |
| 16 | Schougaard et al., 2012 | Indoor pedestrian navigation based on hybrid route planning and location modeling. |
| 17 | Rehrl et al., 2014 | Field study evaluating pedestrian navigation with augmented reality, voice instructions, and a digital map. |
| 18 | Rovelo et al., 2015 | Studying the user experience with a multimodal pedestrian navigation assistant. |
| 19 | Low & Lee, 2015 | Interactive indoor navigation system using visual recognition and pedestrian dead reckoning techniques. |
| 20 | Amirian & Basiri, 2016 | Landmark-based pedestrian navigation using augmented reality and machine learning. |
| 21 | Brata & Liang, 2020 | Comparative study of user experience between a digital map interface and location-based augmented reality. |
| 22 | Tang & Zhou, 2020 | Usability assessment of an augmented reality-based pedestrian navigation aid. |

|    | **Paper** | **Topic of Study** |
|----|-----------|--------------------|
| 23 | Anbaroglu et al., 2020 | Comparison of augmented reality vs. a paper map for pedestrian wayfinding outcomes. |
| 24 | Ng & Lim, 2020 | Design of a mobile augmented reality-based indoor navigation system. |
| 25 | Lee et al., 2020 | Use of AR to help in locating specific books in a library environment. |
| 26 | Kamalam et al., 2022 | Augmented reality navigation for wearable devices with machine-learning techniques. |
| 27 | Preethae al., 2023 | Design and implementation of an augmented reality mobile application for finding bank ATMs. |
| 28 | Gerstweiler et al., 2018 | Development and testing of a dynamic AR guiding system for indoor environments. |
| 29 | Nizam et al., 2021 | Indoor navigation support for the student halls of residence using augmented reality. |
| 30 | Kasprzak et al., 2013 | Evaluation of feature-based indoor navigation using augmented reality. |
| 31 | Rubio-Sandoval et al., 2021 | An indoor navigation method for mobile devices by integrating augmented reality with the Semantic Web. |
| 32 | Calle-Bustos et al., 2021 | Evaluating visual vs. auditory cues in AR for indoor guidance. |
| 33 | Zhang & Nakajima, 2020 | Gamified AR navigational system to prompt residents to explore new areas of a city. |
| 34 | Joo-Nagata et al., 2017 | Augmented reality and pedestrian navigation in the context of an educational program in Chile. |
| 35 | Montuwy et al., 2019 | Effectiveness and UX for older adults using sensory wearable devices to navigate an urban environment. |
| 36 | H. S. Huang et al., 2012 | Spatial knowledge acquisition with mobile maps, augmented reality, and voice instructions. |
| 37 | San Martin & Kildal, 2021 | Audio-visual mixed reality representation of hazard zones for safe pedestrian navigation. |
| 38 | Chaturvedi et al., 2019 | Evaluating a novel AR application for enhancing pedestrian's peripheral vision. |
| 39 | Ajmi et al., 2019 | Development and evaluation of an AR system to assist people with motor disabilities. |
| 40 | Makimura et al., 2019 | Visual effects of turning point and travel direction for outdoor navigation using a head-mounted display. |
| 41 | Rehman & Cao, 2017 | Comparing handheld AR devices versus Google Glass for indoor navigation. |
| 42 | McKendrick et al., 2016 | Neuroergonomic differentiation of hand-held and AR wearable displays during outdoor navigation. |
| 43 | Lee, 2022 | Benefit analysis of a gamified augmented reality navigation system. |
| 44 | Debandi et al., 2018a | Enhancing cultural tourism by a mixed reality application for outdoor navigation and information browsing. |
| 45 | Lu et al., 2021 | Use of AR for enhancing navigation in a university campus environment. |

|    | **Paper** | **Topic of Study** |
|----|-----------|--------------------|
| 46 | Araujo et al., 2019 | Improving fluid transitions between environments in mobile augmented reality applications. |
| 47 | Nenna et al., 2021 | Using AR to investigate cognitive–motor tasks during outdoor navigation. |
| 48 | Dunser et al., 2012 | Evaluating the use of handheld AR during outdoor navigation. |
| 49 | Liu et al., 2021 | Spatial knowledge acquisition with virtual semantic landmarks in mixed reality-based indoor navigation. |
| 50 | Kuwahara et al., 2019 | Evaluation of a campus navigation application using an AR character guide. |
| 51 | Torres-Sospedra et al., 2015 | Enhancing integrated indoor/outdoor mobility on a "smart campus." |
| 52 | Huang et al., 2019 | An augmented reality sign-reading assistant for users with reduced vision. |
| 53 | Munoz-Montoya et al., 2019 | Evaluating the use of AR to assist in object location, including the role of gender. |
| 54 | Sekhavat & Parsons, 2018 | The effect of different AR tracking techniques on the quality of user experience. |
| 55 | Zhao et al., 2020 | The effectiveness of visual and audio wayfinding guidance on smartglasses for people with low vision. |
| 56 | Truong-Allie et al., 2021 | Use of adaptive AR guidance for wayfinding and task completion. |
| 57 | Goldiez et al., 2007 | Effects of augmented reality display settings on human wayfinding performance. |
| 58 | Dong et al., 2021 | Evaluating augmented reality vs. 2D navigation for pedestrian wayfinding. |
| 59 | Zhang et al., 2021 | Enhancing human indoor cognitive map development and wayfinding performance with AR navigation systems. |
| 60 | Mulloni et al., 2011 | User experiences of AR navigational support delivered via smart-phones. |
| 61 | Xie et al., 2022 | Using AR maps to support volunteers in navigating unfamiliar environments. |
| 62 | Mulloni et al., 2012 | Indoor navigation with mixed reality world-in-miniature views and sparse localization on mobile devices. |
| 63 | Mulloni et al., 2011 | Handheld augmented reality indoor navigation with activity-based instructions. |
| 64 | Kerr et al., 2011 | User experiences of wearable mobile augmented reality for outdoor navigation. |
| 65 | Möller et al., 2014 | Evaluation of different types of AR interfaces for visual indoor navigation. |

**3.3. Quality Assessment**

To evaluate the quality of the included articles, we used the Mixed Methods Appraisal Tool (MMAT) within the Covidence platform (Hong et al., 2018). This tool provides a detailed checklist to assess articles used in systematic reviews, taking into account the type of research conducted in each article. Based on our evaluation, 9 of the articles were found to be qualitative research, 10 were found to be quantitative research, and 46 were mixed methods studies. To ensure consistency, two researchers each reviewed all articles separately and then discussed the interpretation and finalized quality score for each study as a group. The quality score calculations were conducted for these categories according to the MMAT and as discussed by Evangelio and colleagues (2022). Each item was scored as Yes (1), Cannot tell (0.5), and No (0). Each article in qualitative and quantitative categories had a total score of 7 (2 screening questions and 1 section), while the mixed methods one had 17 (2 screening questions and 3 sections), hence scores of studies in the former categories were scaled up to 17. The results showed that 19 studies were scored below 7 while the rest were well above 10. Although the overall quality of the included studies was sound, several methodological issues were common across studies with lower quality. Most studies failed to consider potential demographic differences in design or analysis (n=20). Less common methodological issues included: failing to provide clear research questions (n=15) and/or did not address research questions with collected data, failing to interpret findings adequately from experimental data (n=15), using custom designed measures not suitable and/or failing to provide reliability data of these measures for experiment outcomes (n=12), and failing to integrate results from qualitative and quantitative components (n=11). ....Appendix A shows a summary of this assessment.

### 3.4. Data Extraction and Analysis

We created a standardized form for summarizing the articles' contents, and then divided the 65 articles among the two researchers for review. In addition to recording the year of publication, authors' names, type of publication, and the name of the publication outlet, we summarized each study's content in the following categories: (a) research questions and hypotheses; (b) environmental settings or context; (c) country in which the study took place; (d) specific AR technology used; (e) experiment design; (f) independent, dependent, and mediating variables; (g) number of participants and demographic breakdowns (including age and any medical conditions studied); (h) summary of results; (i) limitations and recommendations; and (j) other pertinent information. For most of these categories, we established predefined codes that could be readily applied, while remaining open to studies that might not fit within our a-priori schema. For example, under the category of experiment design we included codes such as between-subjects, within-subjects, or factorial design. Similarly, to understand the studies' results, we identified whether the studies were oriented towards technological advancements, user experience, wayfinding performance, or spatial cognition. In addition to applying these codes we noted the particulars of the results and any notable findings. Finally, we also recorded the number of citations for each article according to Google Scholar as of June 15, 2023. After summarizing all of the studies in this fashion, we carefully reviewed and discussed the collected literature to identify emerging areas of consensus and contradiction.

### 4. Results

The literature review provided robust answers to our four research questions, as presented in the following sections.

## 4.1. The Current State-of-the-art of AR Navigational Assistance Technologies (RQ1)

AR navigation system has applied a variety of technologies to superimpose real-time, context-aware information onto the user's field of view. Key components of the system that were frequently addressed in the reviewed literature include the display device, localization methods for precise user positioning, and the interface design.

***Display Devices.*** Smartphones and tablets have become prominent platforms for AR displays due to their ubiquity and inherent technological capabilities. Among the papers we reviewed, 41 used smartphones and 2 used tablets (Figure 2). Since today's smartphones are almost universally equipped with high-resolution screens, powerful processing units, and inbuilt GPS, accelerometers, and gyroscopes that provide accurate tracking and positioning, it is unsurprising that such devices played a prominent role in the research literature. However, in recent years transparent HMDs ("smart glasses") have become more widely available and are beginning to occupy a larger share of AR projects. These displays provide a hands-free experience by incorporating sensors and cameras directly in the headset for environmental scanning and user interaction. They can also provide sophisticated audio components to enhance the user's sense of immersion. The papers that we reviewed included 12 using Microsoft HoloLens, 3 using Google Glasses, and 6 using custom-designed HMDs. This latter group is particularly notable, as early adopters such as Kerr and colleagues (2011) designed their own head-mounted systems, a process that helped to spur the technology's development and contributed to today's more sophisticated smart-glasses systems (Makimura et al., 2019; Truong-Allie et al., 2021).

**Figure 2.** Types of AR Display Devices Used in the Reviewed Literature

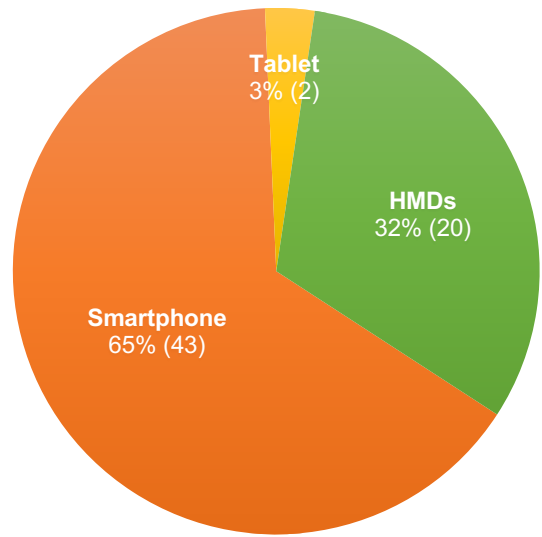

***Localization Techniques.*** The importance of localizing and tracking techniques in AR pedestrian navigation applications cannot be overstated, as they provide the precise positioning necessary to successfully maintain alignment between the virtual components and the physical environment. In our review, all of the studies that were focused on outdoor navigation (n=15) used GPS as the exclusive tracking technique. The indoor studies exhibited more diversity in their approaches (Figure 3). A marker-based technique was featured in 20 articles, in which QR codes or other location markers were placed within the environment and recognized by a camera on the user's mobile device. Mulloni and colleagues' (2011) study exemplified this technique, using floor-mounted fiduciary markers at selected indoor locations. Upon detecting a marker, the app would automatically update its action-based instructions, providing current location, map orientation, and directions to the destination. A more recent study by Rubio-Sandoval and colleagues (2021) combined QR code scanning and pose tracking, the former to establish the users' general position and the latter for fine-grained user location updates within a Unity3D coordinate system.

SLAM was used in 22 of the reviewed studies. In this approach, maps are created and updated using a computational method that also tracks the location of an agent (Durrant-Whyte & Bailey, 2006). The SLAM studies were further divided into 12 that used a computer-vision approach, 6 that used the features of the Microsoft HoloLens, and 4 that used other scanning approaches such as Light Detection and Ranging (LiDAR). The computer vision approach employs cameras and image processing techniques to track feature points across frames for localization. One prominent study in this category was conducted by Huang and colleagues (2019), who used optical character recognition for text-based sign identification. Among the novel localization approaches found in the reviewed literature was a study by Xie and colleagues (2022), in which the researchers built a 3D environmental map using a LiDAR scanner and then employed the iOS's ARKit features to display the user's location and orientation within this map.

**Figure 3.** Types of Localization Techniques Used in the Reviewed Literature

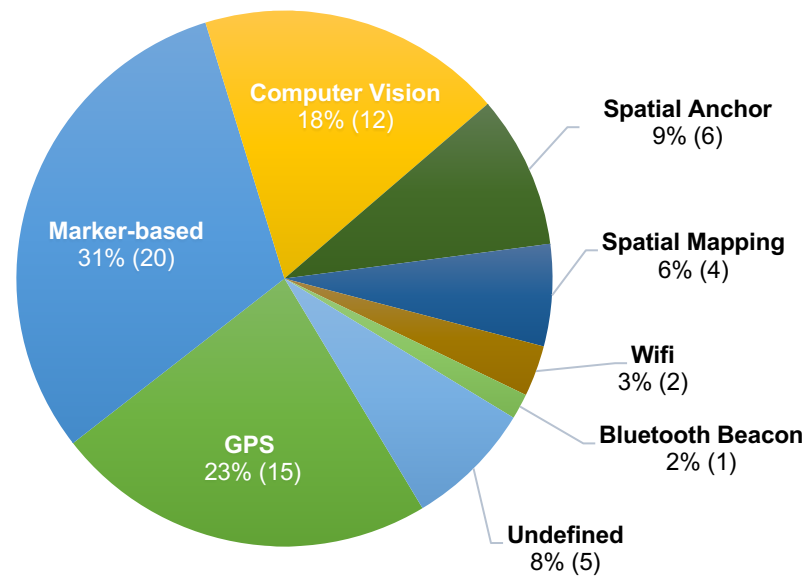

***Pathfinding Techniques.*** Creating paths is critical for rendering navigational aids in AR by determining the optimal routes. Out of 46 articles clearly identified pathfinding techniques, 28 depended on predetermined paths for navigation tasks. This prevalence could be attributed to the specific research objectives requiring participants to adhere to particular wayfinding tasks.

For instance, Montuwy and colleagues (2019) used AR glasses to present predefined, turn-by-turn directions at intersections to navigate users, whereas Kim and colleagues (2015) employed markers to signify turn points, guiding users from one marker to the next. Conversely, 18 articles reviewed employed real-time path generation algorithms. For example, Rubio-Sandoval and colleagues (2021) applied Dijkstra's algorithm to determine the shortest path, considering adjacency and array of nodes in matrices that represented the spatial model. Gerstweiler and colleagues (2017) introduced the FOVPath algorithm, which initially used the A* algorithm to draft a basic path. This was subsequently refined by the Middle Path concept, optimizing navigational guidance based on the user's field of view, allowing for precise path modifications in the immediate surroundings.

***Visualization Techniques and User Interfaces.*** Efficient user interfaces are crucial in AR, representing a significant departure from conventional wayfinding applications that limit users to a two-dimensional map view. When information is being overlayed onto the physical environment it is important not to create too many distractions, so these virtual elements are usually designed in a streamlined fashion. Most of the studies we reviewed combined at least two visualization techniques, including path visualization (26 studies), the use of an auxiliary map (20 studies), displays of distance to a destination or turning point (19 studies), and action-based instructions or arrows (18 studies) (Figure 4).

Dong and colleagues (2021), for example, used an AR module to overlay a blue-colored path onto the camera view, alongside screen-fixed turn arrows indicating the path to the destination. Kim and colleagues (2015) proposed a system featuring multiple real-time-updated arrows to indicate direction and distance to destinations on a smartphone, with additional destination-related information provided via floating virtual tags. Some researchers enhanced the navigation process by overlaying a 2D map onto the path and supplementing it with additional landmark information. For example, the system proposed by Oliveira de Araujo and colleagues (2019) featured an AR browser for outdoor navigation with a hierarchical location display, a simple point-of-interest (POI) search function, a mini radar indicating surrounding POIs, and virtual markers for route guidance. When changing to indoor navigation, their system display adjusted to present location details and building floor plans, while adopting image recognition for localization. Similar multi-functional interfaces were also developed by Sekhavat and Parsons (2018) and Zhang and colleagues (2021).

Some of the studies in our review integrated virtual avatars, interactive quizzes, and other novel visualization techniques into their AR systems to promote user engagement. An example of this is the "BearNavi" app developed by Kuwahara and colleagues (2019), which used a university's bear mascot avatar to guide users around campus while engaging them in educational activities. Similarly, Lee (2022) developed a system in which users followed a virtual AR guide throughout an exhibition, scanning artwork labels for information and amassing virtual souvenirs. Chaturvedi and colleagues (2019) devised a peripheral vision model for smartglasses, which enables users to assimilate information via color and motion detection around the edge of the visual area without shifting their visual focus away from their primary activity.

**Figure 4.** Types of Visualization Techniques Used in the Reviewed Literature

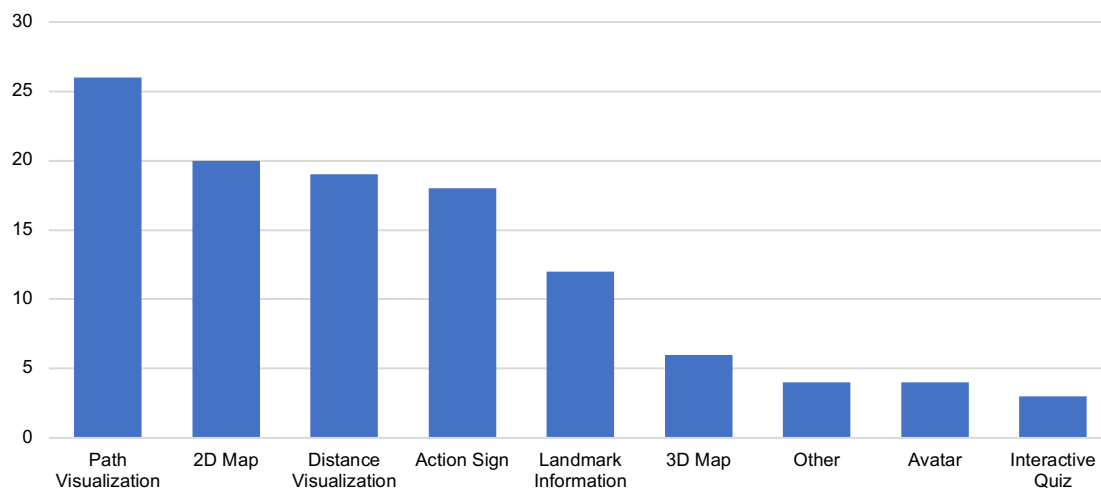

### 4.2. Factors Related to User Experience and the Acceptance of AR Technologies (RQ2)

We identified 49 studies that examined the user experience (UX) of AR navigational assistance systems. These were classified into five distinct categories based on the measurement constructs that they employed (Table 2), including Usability (36 studies), Attitude (21 studies), Perception (13 studies), Acceptance (9 studies), and Engagement (6 studies). Some studies fell into multiple categories and were therefore counted in each. Regarding the display technologies used in these studies, 36 employed handheld devices and 13 employed HMDs. The methods for evaluating UX varied widely across the studies: 31 exclusively used custom questionnaires; 6 employed established system scales such as the System Usability Scale (Brooke, 1986) and the Unified Theory of Acceptance and Use of Technology II Scale (Venkatesh et al., 2012); 3 evaluated user experience based on interviews; 5 employed a combination of custom questionnaires and interviews; and the remaining 4 combined custom questionnaires with existing system scales.

*Usability.* We defined usability in the context of AR navigation systems as the degree to which the technology enabled users to complete wayfinding tasks effectively, efficiently, and satisfactorily (International Organization for Standardization, 1998). Out of the 36 reviewed studies that focused on usability, 27 used comprehensive metrics to examine the overall usability, while the rest honed in on specific aspects such as ease of use, efficiency, learnability, and comfort. The majority of the studies reported that participants gave high usability ratings to AR navigational support systems. However, there were a few notable papers that highlighted usability problems. Tang and Zhou (2020) evaluated the usefulness of an AR outdoor navigation app for both older and younger adults, and found several issues including discrepancies between the included virtual map and the real environment, confusion about the meaning of instructions, abrupt changes in routes, and inaccurate tracking. Zhang and colleagues (2021) reported that while participants gave a HoloLens-based navigational assistance app high scores for effectiveness, they also found the technology difficult to use, particularly when trying to scan for QR code-based markers. Therefore, while AR navigation systems show promising usability overall, the specific design of these systems needs to be carefully evaluated to identify and overcome any emerging usability challenges for the targeted population.

*Acceptance.* Only 9 studies in our review evaluated metrics of technology acceptance, which refers to the willingness of participants to adopt AR for regular use. Of these, 6 examined overall acceptance, while 3 evaluated various constellations of related constructs such as perceived usefulness and learnability. The findings of all studies in this area were consistently positive. Kasprzak and colleagues (2013), for example, demonstrated that participants preferred to use an AR-based navigation support system for locating destinations, compared to seeking out real-world information points or asking other people. Similarly, Debandi and colleagues (2018) developed an application that provided navigational support along with additional information about historical buildings through an HMD. Their study found that users were highly accepting of the technology and viewed it as beneficial. Dunser and colleagues (2012) compared user acceptance between a handheld AR device, a digital map, and a combined condition involving both AR and a digital map. They found that users rated the AR condition highly but preferred the combined interface.

*Attitude.* Attitude towards AR navigation support systems, which is defined as a user's positive or negative feelings about interacting with the technology (Fishbein & Ajzen, 1975), was examined in 21 studies. Our review indicated that participants tended to exhibit a strong preference for AR navigation over traditional methods, and that they tended to favor visual AR modalities over auditory information. For example, Calle-Bustos and colleagues (2021) found that participants reported more positive attitudes toward visual AR feedback compared to audio, and that they cited enhanced entertainment value and reduced attention requirements when explaining this preference. Lu and colleagues (2021) found that participants had more positive attitudes toward immersive 3D map displays compared to 2D AR maps. When compared against the use of physical signs for navigation, AR consistency elicited more positive attitudes—for example, Huang and colleagues (2019) created a sign-reading app capable of identifying real-world text and highlighting it in AR, and found that users had more positive attitudes toward the AR navigation when it was compared directly against non-aided navigation experiences.

    *Perception.* Perception in the realm of AR refers to a user's subjective assessment of how well they can see virtual navigation instructions such as turn-arrows and virtual markers, as well as their simultaneous awareness of situational information in the real-world environment. Of the 13 studies in our review that measured related constructs, 5 explored overall perception, 5 focused on the understanding of virtual cues, and 3 evaluated situational awareness. The vast majority of this literature reported that participants had a clear perception of the AR and physical elements. Sekhavat and Parsons (2018), for example, developed an AR system that provided both navigational instructions and point-of-interest textual descriptions, and found that users had no difficulties perceiving and understanding both types of information. Intriguingly, Kuwahara and colleagues (2019) studied user perceptions of an AR avatar used for route guidance, and discovered that the inclusion of this character increased user's reported perception of environmental information. However, Montuwy and colleagues (2019) noted that some participants struggled to perceive navigational instructions provided in a smartglasses app, and concluded that the weak intensity and short duration of the guidance contributed to these problems.

    *Engagement.* There only a few studies (n=6) in our review that evaluated levels of engagement, immersion, and/or presence. The reasons for this dearth in the literature are not entirely clear, but it may be the case that these constructs are associated more with fully artificial VR environments and less often occurs to AR researchers to consider them. One notable related study in our review was conducted by Ping and colleagues (2020), who designed an outdoor

navigational support system for a cultural heritage site that also enabled users to interact with virtual historical characters and engage in puzzle games associated with various POIs. When evaluating this AR system, the researchers found that participants who used it reported higher engagement and immersion levels and had greater subsequent content understanding compared to a non-AR group. Similarly, Zhang and Nakajima (2020) created a gamified navigation guide for city exploration that included interactive storyline activities at POIs. Participants in the study reported that the interactive elements increased their sense of engagement with the destinations and enriched their recollections of the environment.

**Table 2.** Constructs Employed to Evaluate User Experience in the Reviewed Literature

| Main Categories | Subcategories | Number of Studies per Subcategory | Example References |
|---|---|---|---|
| Usability | Comprehensive (efficiency, effectiveness, satisfaction, ease of use, learnability, usefulness, comfort) | 27 | Mulloni et al., 2011<br>Torres-Sospedra et al., 2015<br>Drewlow et al., 2022<br><br>Click or tap here to enter text. |
| Acceptance | Efficiency | 3 | Xie et al., 2022 |
| | Ease of use | 2 | Möller et al., 2014 |
| | Comfort | 2 | Truong-Allie et al., 2021 |
| | Learnability | 1 | Ping et al., 2020 |
| | Perceived usefulness | 3 | Debandi et al., 2018<br>Montuwy et al., 2019 |
| | Comprehensive (perceived ease of use, perceived usefulness) | 6 | Arntz et al., 2020<br>Zhang et al., 2021 |
| Attitude | Preference | 12 | Anbaroglu et al., 2020<br>Dunser et al., 2012<br>Zhao et al., 2020 |
| | Comprehensive (preference, frustration, trust, confidence, aesthetics, pleasantness) | 2 | Lu et al., 2021 |
| Perception | Frustration | 3 | Kasprzak et al., 2013 |
| | Confidence | 3 | Joo-Nagata et al., 2017 |
| | Pleasantness | 1 | San Martin & Kildal, 2021 |
| | Understanding of virtual guidance | 5 | Möller et al., 2014<br>Oliveira de Araujo et al., 2019 |

| Main Categories | Subcategories | Number of Studies per Subcategory | Example References |
|---|---|---|---|
| | Comprehensive (understanding of guidance, situation awareness) | 5 | Kuwahara et al., 2019 |
| Engagement | Situation awareness | 3 | Sekhavat & Parsons, 2018 |
| | Interactivity | 3 | Kerr et al., 2011 Zhang & Nakajima, 2020 |
| | Immersion | 1 | Hou & Tang, 2020 |
| | Comprehensive (interactivity, immersiveness) | 1 | Ping et al., 2020 |

## 4.3. Effects of AR Navigational Assistance on Human Wayfinding Performance (RQ 3)

A total of 29 empirical studies were found that used comparative measurements of wayfinding performance. Of these studies, 15 compared AR against non-AR conditions, while 14 compared between different AR modalities or evaluated other potentially moderating variables such as gender (Table 3). The experiment settings also varied, with 17 studies conducted in indoor environments, 11 in outdoor environments, and 1 using both indoor and outdoor contexts. In relation to the AR display method used, 19 of these studies employed handheld devices (smartphones and/or tablets), and 10 employed HMDs. The objective measures of wayfinding performance most frequently used were the task completion time and the number of errors committed. Additional domain-specific measures were also represented in various studies, such as the number of stops/pauses, the number of steps taken, the length of deviation from shortest path, and the participants' walking speed.

Notably, ten (67%) of the studies that compared AR-supported navigation against traditional guidance methods, such as paper maps or existing physical signage, observed superior wayfinding performance in the AR condition, as measured by reduced task completion times and/or fewer navigational errors. For example, Zhang and colleagues (2021) tasked participants with indoor wayfinding tasks either using an AR navigation-support system through the HoloLens, or without such assistance. The findings indicated that the AR assistance significantly reduced average wayfinding time, incorrect path length, number of incorrect decisions, and number of pauses. Similarly, Rehman and Cao (2017) reported a significant reduction in task completion times in a complex indoor environment when using handheld AR devices as compared to paper maps. The five studies in our review that did not observe superior wayfinding performance when using AR were all inconclusive; that is, they found no significant differences in the AR vs. non-AR conditions. The reasons for this are unclear and it may be that the data was not extensive enough to confirm fine-grained statistical differences. Rehrl and colleagues (2014) argued, however, that the AR navigational system that they evaluated might be inferior to a GPS-enhanced digital map and to a voice-only guidance system.

Four studies in our review used performance measures to evaluate the integration of interactive maps into AR systems. Generally, there was a positive correlation between the adoption of map interfaces and improved wayfinding performance. In a study conducted by Hou and Tang (2020), the integration of a 2D layout map into the AR interface resulted in participants completing navigation tasks faster and with fewer steps taken. Mulloni and colleagues (2011) combined action-based instructions with a 3D layout map at selected information points in an indoor environment, and found that the addition of the map enhanced wayfinding task performance. Conversely, however, Tang and Zhou (2020) found no significant differences in navigation errors and walking speed when adding 2D maps to an AR interface alongside path visualizations. Xie and colleagues (2022) compared the use 3D vs. 2D maps in an indoor AR system for visually impaired participants, and found that using the 3D maps reduced wayfinding task completion time.

An additional four studies in the review compared different AR guidance modalities, including visual, auditory, and tactile. Three of these studies (75%) found visual guidance to be superior at enhancing wayfinding performance, with the final study finding no significant differences. Zhao and colleagues (2020) developed a visual and audio wayfinding guidance system using an HMD, and found that participants made fewer errors and had faster task completion time when using the visual guidance, even for some participants who had partial vision impairments. Similarly, Calle-Bustos and colleagues (2021) reported more deviation steps take by participants when using auditory guidance on smartphones compared to visual guidance. Rovelo and colleagues (2015) found improved wayfinding performance when using visual feedback via an HMD compared against audio and tactile feedback (with the latter provided by a smartwatch that vibrated to indicate directional guidance).

Finally, six studies in our review controlled the AR modality while examining moderating or confounding variables affecting performance. One study in this group reported superior wayfinding task completion times and less distance traveled for male participants than female participants when using AR guidance (Ahmad et al., 2005). Two different visual interface designs were compared by Arntz and colleagues (2020), who found that navigational arrows were more effective than path visualizations for improving wayfinding performance. Sekhavat & Parsons (2018) contrasted the efficacy of a marker-based localization method that utilized QR code scanning for navigation guidance, with a location-based approach that used GPS, allowing participants to explore Points of Interests (POIs) outdoors. The findings indicated that the location-based method resulted in significantly less task completion time and fewer navigation errors.

**Table 3.** Studies in the Review that Compared Wayfinding Performance

| Framework | Study Result | Reference | Comparison | Dependent Measures | Setting |
|---|---|---|---|---|---|
| AR vs. Non-AR | AR higher performance | Anbaroglu et al., 2020 | Handheld to paper map | Task completion time | Outdoor |
| | | Goldiez et al., 2007 | HMD to paper map | Task completion time | Indoor |

| Framework | Study Result | Reference | Comparison | Dependent Measures | Setting |
|---|---|---|---|---|---|
| | | Huang et al., 2019 | AR signs to no guidance | Extra distance ratio, walking speed, task duration | Indoor |
| | | Kasprzak et al., 2013 | Handheld to paper map | Task completion time, wrong turns and pauses | Indoor |
| | | Montuwy et al., 2019 | HMD to digital map | Time to destination, number of errors | Outdoor |
| | | Lee et al., 2020 | Handheld to text signs | Task completion time, distance travelled | Indoor |
| | | Rehman & Cao, 2017 | HMD and handheld to paper map | Task completion time, route retention error | Indoor |
| | | Rubio-Sandoval et al., 2021 | Handheld to no guidance | Navigation time | Indoor |
| | | Smith et al., 2017 | AR to no guidance | Number of waypoint decisions made | Outdoor |
| | | Zhang et al., 2021 | HMD to no guidance | Wayfinding time, extra path length, number of incorrect decisions, number of pauses | Indoor |
| | No difference | Dong et al., 2021 | Handheld to digital map | Wayfinding duration, frequency of mistakes or hesitation events | Outdoor |
| | | Dunser et al., 2012 | Handheld to digital map | Time to navigate between locations, distance travelled | Outdoor |
| | | Huang et al., 2012 | Handheld to digital map and voice guidance | Completion time, number of stops | Outdoor |
| | | Lee, 2022 | Handheld to 2D map | Distance travelled, travel time | Outdoor |
| | | Rehrl et al., 2014 | Handheld to digital map and voice guidance | Completion time, number and duration of stops | Outdoor |
| AR vs. AR | Map higher performance | Hou & Tang, 2020 | 2D map to no map | Task completion time, step deviation | Indoor |
| | | Mulloni et al., 2011 | 3D map to no map | Step difference, task completion time, navigation errors | Indoor |
| | | Xie et al., 2022 | 3D map to 2D map | Task completion time | Indoor |

| Framework | Study Result | Reference | Comparison | Dependent Measures | Setting |
|---|---|---|---|---|---|
| | No difference for map | Tang & Zhou, 2020 | 2D map to no map | Success rate, navigation error, speed variation | Outdoor |
| | Visual higher performance | Calle-Bustos et al., 2021 | Visual to audio guidance | Task completion time | Indoor |
| | | Rovelo et al., 2015 | Visual to tactile guidance | Task completion time, number of errors | Outdoor |
| | | Zhao et al., 2020 | Visual to audio guidance | Navigation time, error rates | Indoor |
| | No difference for visual | San Martin & Kildal, 2021 | Visual to audio and audiovisual | Task completion time | Indoor |
| Moderating/ Confounding Variables | Less attention load had higher performance | Nenna et al., 2021 | Single task to dual task | Exploration time, walking velocity | Outdoor |
| | Location-based AR had higher performance | Sekhavat & Parsons, 2018 | Location-based AR to marker-based AR | Completion time, error counts | Indoor/ outdoor |
| | Male had higher performance | Ahmad et al., 2005 | Female to male | Task completion time, percentage of maze covered | Indoor |
| | Minimal UI had higher performance | Truong-Allie et al., 2021 | Minimal UI to complex UI | Task completion time | Indoor |
| | Navigation path had higher performance | Arntz et al., 2020 | Navigation path to arrow guidance | Path length, path time | Indoor |
| | AR video had higher performance | Möller et al., 2014 | AR video to handheld | Task completion time | Indoor |

## 4.4. Impacts of AR Navigational Assistance on Human Perception, Decision-making, Behavior, and Cognition (RQ4)

The review included 13 studies that examined cognitive load and cognitive map development and 5 studies that examined aspects of perception when using AR (Table 4). There were no studies found in the review that examined decision-making processes or specific wayfinding behaviors. For those focusing on cognition, 5 studies evaluated cognitive load during navigational tasks; 4 focused on the development of cognitive maps; and 4 studies examined both cognitive load and cognitive maps. Seven of these studies were conducted in indoor areas, and 6 were conducted outdoors. Various measures were employed to gauge the effect of AR on mental workload, with the NASA-TLX (Hart & Staveland, 1988) being the most commonly used instrument. Other studies adopted eye tracking measures such as fixation durations, saccade amplitudes, and pupil size as real-time objective measures of cognitive load (Dong et al., 2021; Kluge & Asche, 2012). One study employed wireless functional near infrared spectroscopy (fNIRS) to non-invasively monitor hemodynamic changes in the brain's prefrontal cortex as a measure of cognitive load (McKendrick et al., 2016). In addition, several studies employed a secondary cognitive task, such as a working memory test, conducted simultaneously with the wayfinding activity, and evaluated cognitive load based on the task scores. Regarding the development of cognitive maps, measurements included sketch map task, object recall task, and custom questionnaires. One study used a post-navigation wayfinding task without aids, after previous AR exploration in the building, to evaluate the establishment of cognitive maps (Zhang et al., 2021).

In 8 out of the 9 studies evaluating cognitive load (89%), the AR condition was associated with a significantly reduced mental workload compared to other navigational approaches. One notable study in this area is Zhao and colleagues (2020), who investigated cognitive load in relation to visual vs. audio AR guidance. Their findings indicated that participants following audio guidance had significantly greater cognitive load—which may help to explain why participants across multiple reviewed studies performed better at wayfinding when they received visual rather than audio guidance (Section 4.3 above). Dong and colleagues (2021) found that participants using AR experienced significantly lower cognitive load than those using a 2D digital map, as evidenced by eye-tracking measures. This study also found that participants using AR directed more attention towards other people in the environment. McKendrick and colleagues (2016) demonstrated that users of head-mounted AR exhibited significantly lower hemodynamic brain responses than those using handheld AR, indicating heightened neural activity and cognitive load in the handheld device group. The single study in our review that found worse outcomes for AR was conducted by Rehrl and colleagues (2014), who observed an increased cognitive load when participants navigated with a handheld AR tool compared to a combination of digital map and voice-based guidance in outdoor settings.

In regard to the development of cognitive maps, 7 out of 8 studies (88%) found that AR navigational aids significantly improved map formation. This was true across multiple types of AR content. For example, Liu and colleagues (2021) implemented a system featuring iconic holograms for virtual semantic landmarks, and subsequently assessed its impact on users' spatial knowledge acquisition. Their results from sketch map and landmark location tasks suggested that such virtual landmarks effectively enhanced route knowledge. Similarly, Zhang and colleagues (2021) demonstrated that the use of virtual guideposts furnished with arrows and labels, and the deployment of 3D interactive layout models in AR, could substantially improve cognitive map development in indoor environments. The outlier study in this area was conducted by Rehman and Cao (2017), who found that despite a reduction in cognitive workload associated with a wearable AR navigation support system, participants who used the AR had more route retention errors compared to those who used paper maps.

The 5 studies in our review that evaluated perception all focused on participants' ability to observe, integrate, and make use of the conveyed virtual information, and all of these studies found that AR enhanced perceptions of the environment. Makimura and colleagues (2019), for example, found that visual cues in a head-mounted display significantly enhanced perception of navigation-related elements in the environment, such as available route choices, compared to a non-AR condition. San Martin and Kildal (2021) similarly found that participants' perception of hazard zones was successfully improved by AR auditory and visual warnings about those zones. The visual information was more effective in improving the accuracy of perceived distances from the user to the hazard, while the auditory feedback was more effective in inciting a perception of danger. Another notable study in this area was conducted by Kluge and Asche (2012), who found that AR users tended to rely more on superimposed environmental information when making route decisions at intersections, while relying more on the included 2D digital map to track their position when not at decision-points.

**Table 4.** Studies in the Review that Addressed Cognition

| Study Result | References | Comparison | Dependent Measures | Setting |
|---|---|---|---|---|
| AR lowers cognitive load | Dunser et al., 2012 | AR with 2D map to AR without map | NASA TLX | Outdoor |
| | McKendrick et al., 2016 | HMD to mobile AR | fNIRS (functional near infrared spectroscopy) | Outdoor |
| | Nenna et al., 2021 | fewer AR objects to more AR objects | NASA TLX, cognitive task | Outdoor |
| | Xie et al., 2022 | 3D map to 2D map | NASA TLX | Indoor |
| AR lowers cognitive load but, inhibits cognitive map formation | Rehman & Cao, 2017 | HMD and handheld AR to paper map | NASA TLX | Indoor |
| AR lowers cognitive load and enhances cognitive map formation | Dong et al., 2021 | handheld AR to paper map | eye tracking, sketch map | Outdoor |
| | Zhang et al., 2021 | HMD to no guidance | NASA TLX, sketch map, questionnaire | Indoor |
| | Zhao et al., 2020 | visual to audio guidance | cognitive task, sketch map | Indoor |
| AR increases cognitive load | Rehrl et al., 2014 | AR system to non-AR audio guidance and mobile map | NASA TLX | Outdoor |
| AR enhances cognitive map formation | Calle-Bustos et al., 2021 | audio to visual | sketch map | Indoor |
| | Huang et al., 2012 | handheld to mobile map | route memory | Outdoor |

| Study Result | References | Comparison | Dependent Measures | Setting |
|---|---|---|---|---|
| | Liu et al., 2021b | HMD only | sketch map | Indoor |
| | Munoz-Montoya et al., 2019 | handheld only | object recall task, spatial anxiety | Indoor |
| AR increases perception accuracy | Anbaroglu et al., 2020 | handheld to mobile map | questionnaire | Outdoor |
| | Kluge & Asche, 2012 | handheld only | eye tracking, questionnaire | Outdoor |
| | Makimura et al., 2019 | visual guidance only | questionnaire | Outdoor |
| | Möller et al., 2014 | video to handheld | questionnaire | Indoor |
| | San Martin & Kildal, 2021 | visual to audio | NASA TLX, interview | Indoor |

## 5. Discussion

The current state-of-the-art of AR navigational support technology as revealed in the literature review indicates that no single type of display device, localization method, or visualization strategy has yet achieved hegemony. Smartphones have remained one of the most common platforms for delivering AR content since its commercial debut, but head-mounted displays also account for a substantial portion of AR navigation research. The presentation formats of these two device types are substantially different, with impacts for users' overall body positioning and freedom of movement. Thus, it is important to account for the specific device that is used when evaluating and applying research findings. At the same time, user-localization techniques differ among various studies, particularly in indoor environments. Marker-based approaches and the application of GPS for pinpointing users' location continue to be implemented, but the field appears to be moving toward the more pervasive use of computer vision and image recognition to help precisely overlay virtual elements (Delgado et al., 2020). Advanced HMDs such as the HoloLens have incorporated internal depth sensors and cameras for improved localization, and some recent studies have branched out into using LiDAR or lasers to perform environmental scanning. In addition to the transition towards integrated hardware solutions, there has been a noticeable move towards unified software solutions, utilizing third-party platforms and SDKs to enable developers to create applications for various hardware types and deploy them across multiple operating systems. Notably, Unity (Unity Technologies, 2023) has been widely used for application development, with SDKs like ARKit (Apple Inc., 2023) and ARCore (Google LLC., 2023) being employed for mobile platforms, and Mixed Reality Toolkit (Microsoft Corporation, 2023) for HMDs. As these techniques continued to advance, there may be opportunities to further integrate multiple localization methods into a single technology, which could enhance spatial transitions and make the AR system more flexible for use in diverse environments.

Visualization techniques and user interfaces have seen equally dynamic growth. The increased incorporation of multimodal interactions, including visual, auditory, and tactile cues, has provided designers and users with more options for information delivery. Visual cues have traditionally comprised turn-by-turn instructions such as path visualization, turn indicator arrows, distance indicators, action signs, and auxiliary 2D maps. However, the integration of landmark information with interactive functionalities, peripheral vision models, avatars that engage the user conversationally, and other novel techniques has led to a great diversification in AR presentations. There is still very little empirical research on the effectiveness of some of these modalities.

Our evaluation of literature related to the user-experience aspects of AR navigation systems indicated that there was a great deal of variance across individual modalities. Simple, streamlined visual presentations, with effective localization, tended to receive the highest user praise. Some studies identified specific challenges for UX, which were mostly related to platform design, such as discrepancies between virtual and real routes, abrupt changes in guidance, requirements for users to constantly scan the environment to look for markers, and overly complex or confusing display elements. We recommend that AR system designers should carefully study this literature to become aware of potential pitfalls, and that new systems should be carefully user-tested to check for emerging problems. Among the most consistent findings in our review was that users disliked systems that relied purely on auditory information signals, and that they approved of systems that visually highlighted physical features in the environment salient to wayfinding. It should be noted that, despite the occasional negative findings, participants in general in the reviewed studies found the AR wayfinding systems to be highly useful and enjoyable. Preference for AR over traditional methods such as digital or paper maps was often linked to interactivity, engagement, and entertainment value—factors that designers can leverage to improve the user experience and acceptance of the technology. Incorporating diverse modalities and personalization options so that users can tailor the experience to their personal needs and preferences will likely further enhance this positive reception.

When it comes to the effects of AR on wayfinding performance, findings in the literature were mixed. This was particularly the case when contrasting indoor vs. outdoor contexts. In indoor environments, AR consistently outperformed conventional methods such as paper or 2D digital maps, reducing task completion times and navigational errors. It was more common in outdoor navigation for studies to find no substantial differences between AR and conventional guidance. This outcome could potentially indicate wayfinders were using different cues or navigational strategies in outdoor vs. indoor contexts, and that the guidance provided by AR was more effective for indoor use. However, other factors, such as the intensity of the outdoor lighting affecting users' perception, may also play a role. Another highly notable finding in this area that all of the studies that found no positive impact of AR were conducted using handheld devices. We suggest that the requirement for constant physical engagement in interacting with the handheld platforms might be a contributing factor that detracts from the effectiveness of AR, especially in outdoor contexts.

We found that visual guidance typically outperformed auditory guidance. Tactile guidance provided the least wayfinding performance benefits, but it was only evaluated in a single study, and thus more research is needed to draw any conclusions about this novel modality. The inclusion of interactive maps as part of the AR experience was consistently shown to improve wayfinding performance compared to purely directional guidance, However, users relied on the directional guidance more often than maps at important decision-making points, highlighting the centrality of overlayed path visualizations. Overall, the evidence collectively indicates that the effect of AR on wayfinding performance is highly context-dependent, influenced by variables such as gender, engagement level, device used, accuracy in localization, visibility of navigational cues, type of environment, and design considerations. Unfortunately, most of the existing literature has considered only one or two of these variables at a time, making it difficult to draw broader conclusions across disparate studies. Researchers should continue to pursue comparative studies that will directly juxtapose multiple variables within the same experiment framework, for example by using different device types and multiple feedback strategies across both indoor and outdoor routes.

Concerning the cognitive effects of using AR navigational tools, a central finding of the review is that a large body of literature employing varied measurement methods has demonstrated that AR reduced mental workload in comparison to non-AR wayfinding. This was manifested in multiple ways, including assessment surveys such as the NASA-TLX and eye tracking measures such as fixation durations. Generally, participants using AR through head-mounted displays reported lower cognitive load than mobile devices, and both were found to be superior to paper maps or unguided navigation. Additionally, visual guidance reported a lower cognitive load than audio guidance. These findings about cognitive load strongly overlap with the wayfinding performance outcomes and with the user preferences regarding these modalities. The results are consistent with Cognitive Load Theory as applied to AR, which holds that the technology's in-context presentation of information can reduce the mental effort required to understand and assimilate navigational guidance. By off-loading the mental task of processing and visualizing navigation guidance to real-world overlays, AR presents a more streamlined and less distracting experience (Fan et al., 2020; Kim & Dey, 2016).

The majority of the reviewed studies found that AR significantly aided cognitive map development, compared to conventional forms of navigational guidance. This is again consistent with the split-attention effect from the Cognitive Load Theory, which holds that the reduction in effort obtained from direct and immersive presentation of spatial information will leave more cognitive resources available for the robust formation of maps (Goff et al., 2018). Nevertheless, one study in our review (Rehman and Cao, 2017) found that reduced cognitive workload from using AR did not translate into better cognitive maps, with the AR group actually scoring significant worse in terms of route memory errors, possibly due to the deficiencies of digital navigation guidance in aiding user understanding of the routes. This finding appears to be idiosyncratic, but it is an important caution suggesting that more research is needed to investigate how different AR devices, information modalities, and environmental contexts may intersect to influence spatial knowledge acquisition. Moreover, current AR solutions predominantly offer generalized wayfinding guidance without catering to individual needs such as preferences and expertise levels, potentially leading to overreliance on technology and stifle spatial learning. More studies are essential to devise scaffolding mechanisms that tailor AR guidance to individual's cognitive processes (Moghaddam et al., 2021).

**6. Limitations and Suggestions for Future Research**

Methodologically, this systematic literature review is constrained by its selection criteria; for example, it does not address AR-supported navigation while driving an automobile or navigation within fully virtual environments. The exclusion of studies not written in English may have resulted in some relevant articles being overlooked. It should also be noted that the number of included studies was somewhat thin in certain areas of our research interest. User experience issues were discussed in 49 of the included studies, but there were only 29 studies in the review that addressed wayfinding performance, and only 18 that addressed spatial cognition. We did not find any studies at all addressing AR's relation to navigational decision-making processes or specific wayfinding behaviors. Moreover, many of the studies in the review suffered from small sample sizes, often including fewer than 10 participants. This scarcity of data may limit their ability to derive statistically significant insights. It is not entirely surprising to see a dearth of research literature in these areas since AR is a relatively new technology; however, the low number of studies on some topics means that the review's conclusions should be considered provisional until further research can be conducted.

Additional considerations that may have affected the insights attained during the literature review include the low number of studies considering confounding factors such as participants' gender, age, and geographic background, and the tendency of many included studies to use oversimplified wayfinding tasks (e.g., tasks that do not require a change in floor level). Most of the experiments were quite brief (less than 20 minutes) and therefore may not reflect the user experience of individuals who employ AR devices for extended periods of time. The different types of buildings and unique navigational tasks used in each study can reduce the ability to compare and generalize their findings. Several studies in our review reported technical difficulties such as lagging and inaccurate localization; while these issues seem to improve over time as the technology matures, researchers should carefully pilot-test their approaches to reduce the number of such glitches.

Future research should strive to expand the scope of AR wayfinding studies, to simultaneously consider more variables in the same experiment, to obtain larger and more diverse participant samples, and to use more complex and realistic wayfinding tasks. A central goal should be to delve deeper into specific AR design considerations, and to investigate the nuanced interplay between different information modalities. In the current literature review we had to rely on comparing disparate studies that used distinct wayfinding tasks and protocols to synthesize an overview perspective. Future researchers can build upon the review's conclusions by directly integrating the emerging variables of interest, for example by conducting mixed inside/outside research with both handheld and head-mounted AR displays and various information modalities. This can help to determine if the inside/outside differences noted in the literature review are replicable, and provide insights about the potential underlying mechanisms of such differences. These targeted efforts will help to build a more robust and thorough understanding of the role and potential of AR navigation technology to improve users' wayfinding experiences.

# Appendix A

**Table 1.** Results of MMAT Quality Assessment.

|   | Paper | Assessment Score (Total of 17) |
|---|---|---|
| 1 | Stigall et al., 2019 | 6.7 |
| 2 | Rochadiani et al., 2022 | 6.1 |
| 3 | Ahmad et al., 2005 | 13.5 |
| 4 | Kim et al., 2015 | 14.0 |
| 5 | Smith et al., 2017 | 10.3 |
| 6 | Ping et al., 2020 | 16.5 |
| 7 | Hou & Tang, 2020 | 11.9 |
| 8 | Arntz et al., 2020 | 15.5 |
| 9 | Romli et al., 2020 | 3.8 |
| 10 | Drewlow et al., 2022 | 4.75 |
| 11 | Sheoprashad & Defreitas, 2022 | 6.5 |
| 12 | Wakchaure et al., 2022 | 5.5 |
| 13 | Yunardi et al., 2022 | 6 |
| 14 | Sharin et al., 2023 | 4.75 |
| 15 | Kluge & Asche, 2012 | 8 |
| 16 | Schougaard et al., 2012 | 5.5 |
| 17 | Rehrl et al., 2014 | 14.25 |
| 18 | Rovelo et al., 2015 | 15 |
| 19 | Low & Lee, 2015 | 2.25 |
| 20 | Amirian & Basiri, 2016 | 11.5 |
| 21 | Brata & Liang, 2020 | 6.75 |
| 22 | Tang & Zhou, 2020 | 13.75 |
| 23 | Anbaroglu et al., 2020 | 12.25 |
| 24 | Ng & Lim, 2020 | 6.6 |
| 25 | Lee et al., 2020 | 13.75 |
| 26 | Kamalam et al., 2022 | 3 |
| 27 | Preethaet al., 2023 | 6.25 |
| 28 | Gerstweiler et al., 2018 | 14 |
| 29 | Nizam et al., 2021 | 5 |
| 30 | Kasprzak et al., 2013 | 15 |
| 31 | Rubio-Sandoval et al., 2021 | 13 |
| 32 | Calle-Bustos et al., 2021 | 13.75 |
| 33 | Zhang & Nakajima, 2020 | 5 |
| 34 | Joo-Nagata et al., 2017 | 8.25 |

|   | **Paper** | **Assessment Score (Total of 17)** |
|---|---|---|
| 35 | Montuwy et al., 2019 | 14.75 |
| 36 | H. S. Huang et al., 2012 | 13.25 |
| 37 | San Martin & Kildal, 2021 | 14.75 |
| 38 | Chaturvedi et al., 2019 | 12.5 |
| 39 | Ajmi et al., 2019 | 7.0 |
| 40 | Makimura et al., 2019 | 13.75 |
| 41 | Rehman & Cao, 2017 | 15.5 |
| 42 | McKendrick et al., 2016 | 15.25 |
| 43 | Lee, 2022 | 15.2 |
| 44 | Debandi et al., 2018a | 10.1 |
| 45 | Lu et al., 2021 | 12.25 |
| 46 | Araujo et al., 2019 | 10.9 |
| 47 | Nenna et al., 2021 | 16.25 |
| 48 | Dunser et al., 2012 | 15.5 |
| 49 | Liu et al., 2021 | 14 |
| 50 | Kuwahara et al., 2019 | 10.75 |
| 51 | Torres-Sospedra et al., 2015 | 2.4 |
| 52 | Huang et al., 2019 | 14.25 |
| 53 | Munoz-Montoya et al., 2019 | 8.25 |
| 54 | Sekhavat & Parsons, 2018 | 15 |
| 55 | Zhao et al., 2020 | 15.5 |
| 56 | Truong-Allie et al., 2021 | 15 |
| 57 | Goldiez et al., 2007 | 14.6 |
| 58 | Dong et al., 2021 | 15.5 |
| 59 | Zhang et al., 2021 | 15.5 |
| 60 | Mulloni et al., 2011 | 4 |
| 61 | Xie et al., 2022 | 15.5 |
| 62 | Mulloni et al., 2012 | 11.25 |
| 63 | Mulloni et al., 2011 | 15.1 |
| 64 | Kerr et al., 2011 | 14.5 |
| 65 | Möller et al., 2014 | 16.5 |